\documentstyle[aps,prl,twocolumn,floats,graphicx,amssymb,amsmath]{revtex}

\begin{document}
\draft
\wideabs{
\title{Time-resolved optical observation of spin-wave dynamics}
\author{J.~S.~Dodge, A.~B.~Schumacher\cite{ref:andreasadd},
  J.-Y.~Bigot\cite{ref:jyadd} and D.~S.~Chemla}

\address{Materials Sciences Division, E.~O.~Lawrence Berkeley National
  Laboratory, Berkeley, CA 94720, and\\Department of Physics,
  University of California at Berkeley, Berkeley, CA 94720}

\author{ N.~Ingle and M.~R.~Beasley}

\address{Department of Applied Physics, Stanford University, Stanford,
  CA 94305}

\date{\today}
\maketitle

\begin{abstract}
  We have created a nonequilibrium population of antiferromagnetic
  spin-waves in ${\rm Cr_2O_3}$, and characterized its dynamics, using
  frequency- and time-resolved nonlinear optical spectroscopy of the
  exciton-magnon transition. We observe a time-dependent pump-probe
  line shape, which results from excitation induced renormalization of
  the spin-wave band structure. We present a model that reproduces the
  basic characteristics of the data, in which we postulate the optical
  nonlinearity to be dominated by interactions with long-wavelength
  spin-waves, and the dynamics to be due to spin-wave thermalization.
\end{abstract}

% Time-resolved optical spectroscopies and other ultrafast
%  optical measurements in condensed matter,
% Spin waves
% Antiferromagnetics
% Quantized spin models
}

Optical magnetic excitations have been studied extensively in several
magnetic oxides~\cite{ref:Tanabe1982,ref:Strauss1980}, and have
recently attracted interest in studies of low-dimensional correlated
electron systems~\cite{ref:Lorenzana1995,ref:Damascelli1998}.  In this
letter we demonstrate how time-resolved, nonlinear optical
spectroscopy (TR-NLOS) of optical magnetic excitations may be used to
investigate the interactions and dynamics of short-wavelength magnetic
modes in strongly correlated systems.  These short-wavelength
excitations are the most difficult to treat theoretically, and even in
three dimensions their mutual interaction is not fully understood.  We
present results of femtosecond pump-probe spectroscopy of the
exciton-magnon ($X$-$M$) absorption feature in the antiferromagnetic
oxide ${\rm Cr_{2}O_{3}}$ \cite{ref:Macfarlane1971}, with both
temporal and spectral resolution.  This optical absorption feature
allows us to excite antiferromagnetic spin-waves directly, with
nonequilibrium occupation distributions weighted toward large 
momenta, high energies, and with sufficient density to observe interaction
effects.  We have observed a novel nonlinear optical effect associated
with the nonequilibrium occupation dependence of the spin-wave (SW)
dispersion relation.  In semiconductor physics, it has been
demonstrated that important and nontrivial physics of correlated many-
particle systems can be clarified through careful attention to
dynamics, using TR-NLOS\cite{ref:Chemla98}.  Our work indicates that
TR-NLOS can be used to directly manipulate and study magnetic
excitations in strongly correlated insulators, in addition to the
charged excitations usually probed with optics.

In the periodic lattice of a magnetic crystal the concept of a SW
appears naturally when the fermionic spin-Hamiltonian is expressed in
terms of Boson operators \cite{ref:Mattis1988}.  Controlled
approximations exist for distributions of long-wavelength SWs at low
excitation densities, but away from these conditions theory must be
guided by experimental observations.  Neutron spectroscopy and linear
optical spectroscopy have provided some of the best evidence for the
existence of spin wave renormalization (SWR) at elevated temperatures
\cite{ref:Yelon1971,ref:White1965}, but as previous measurements were
largely limited to thermally occupied SWs, not much is known about the
interactions among excitations at short wavelengths.  Theory suggests
that the interactions may undergo qualitative changes as the zone
boundary is approached~\cite{ref:Wortis1963}.  In our experiments,
using laser excitation of the $X$-$M$ absorption, we are able to
macroscopically occupy a strongly nonthermal distribution of SWs, with
a distribution heavily weighted toward the zone boundary.

The $X$-$M$ transition may be understood as a magnon sideband to an
exciton, and is similar in character to the well-known two-magnon
absorption. In a cubic environment, the three electrons per ${\rm
  Cr^{3+}}$ site possess a ground state with ${\rm ^4A_2}$ symmetry
and a lowest excited ${\rm ^2E}$ state at $\sim1.7$ eV. In a lattice,
these levels couple via superexchange interactions $J$ and
$J^{\prime}$, of order 50 meV. The ground state multiplet
develops into antiferromagnetic-SWs, and the excited state multiplets
into ``magnetic exciton'' bands.  The total spin projection $S_z$ is
preserved by the two spin excitations, so the spin selection rule
$\Delta m = \Delta m_{X} + \Delta m_{M} = 0$ is satisfied. To conserve
momentum the photon is absorbed by an exciton and a magnon of equal
and opposite momentum, ${\bf k}_{X} + {\bf k}_{M}
\sim0$~\cite{ref:Tanabe1965}.  The excitation is largely localized to
neighboring sites, and consequently draws its spectral weight from
states over the entire Brillouin zone.  Neglecting interaction between
the nearby exciton and magnon for clarity, the $X$-$M$ absorption
line shape is given approximately by the joint density of states
(JDOS)~\cite{ref:Macfarlane1971}:
\begin{equation}
    \rho_{e-m}(\omega) = \sum_{\bf k}\delta
    (\omega - \omega_{\bf k}^{e} - \omega_{-\bf k}^{m}).
    \label{eq:jdos}
\end{equation}
The SW dispersion is stronger than that of the exciton, and provides
the dominant contribution to the line shape.  The absorption feature
may be qualitatively understood as the SW DOS, shifted up rigidly in
energy by the exciton energy. This allows us to excite SWs and
subsequently monitor their density of states (DOS), using a pulsed
near-infrared laser.

We have performed pump-probe spectroscopy of the $X$-$M$ transition,
using 100 fs pulses emanating at 76 MHz from a Ti:sapphire laser tuned
to $\hbar \omega \approx$ 1.765 eV. The beam is split in a 10:1
(pump:probe) ratio and the time delay, $\Delta t = t_{probe} -
t_{pump}$ is controlled with a delay line.  Both pump and probe are
focused through a microscope objective to a 6 $\mu$m diameter spot at
the sample, which is held at 10 K with a cold finger cryostat.  We
estimate the average temperature at the sample to be $\le 30$ K. The
sample is a $2.2\ \mu$m thick film of ${\rm Cr_2O_3}$ grown
epitaxially on sapphire, by evaporation of Cr in a reactive oxygen
environment.  The $c$ axis of both ${\rm Cr_2O_3}$ and sapphire
is normal to the film surface.  At the peak of the $X$-$M$ absorption
line, the internal transmissivity of the sample is 52\%.

\begin{figure}[tbp]
     \includegraphics[width=3in]{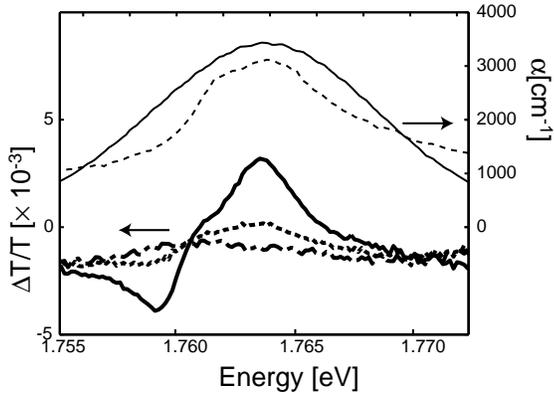}
\caption{Pump-probe spectrum at $\Delta t=$ 0.3 (heavy dash-dotted
  line), 10 (heavy dotted) and 50 ps (heavy solid), with units given
  on the left, shown together with the absorption spectrum (light
  dashed), with units given on the right, and the incident laser
  spectrum (light solid), shown for reference in arbitrary units.}
\label{fig:specrslv}
\end{figure}

After the sample we measure $\Delta{\rm T}\times I_{inc} =
I_{on}-I_{off}$, the change in the transmitted probe intensity as the
pump is chopped mechanically.  In the small signal regime, the
normalized change in transmission reproduces the differential
absorption of the sample, by the relation $\Delta{\rm T/T} \approx -
\Delta \alpha \, L$.  We measure $\Delta{\rm T/T}\simeq(I_{on} -
I_{off})/I_{off}$ as a function of wavelength and time delay, and
substract from this a small background contribution which persists at
negative time delay because of the finite repetition rate.  At
excitation densities of $\sim10^{-3}/{\rm Cr}$, we observe the
complex spectral feature shown in Fig.~\ref {fig:specrslv} for three
different time delays.
The data are well described in terms of two components: a spectrally
featureless photoinduced absorption (PIA), which shifts the overall
$\Delta{\rm T/T}$ toward negative values and is relatively time
independent over 100 ps, and a derivative-like line shape, which is
weakly evident at $\Delta t = 0$ and grows as a spectral unit as
$\Delta t$ increases.

The magnitude and qualitative line shape shown in
Fig.~\ref{fig:specrslv} may be explained by recognizing that the
absorption of each photon creates a SW, which in turn renormalizes the
overall SW band structure through interaction.  In a cubic ferromagnet
with occupation at long wavelengths, this renormalization takes the
form
\begin{eqnarray}
    \omega_{\bf k}(\{n_{\bf k'}\}) & = & \omega_{\bf k}^{0}
    [1 - \frac{1}{zJS^{2}N}\sum_{\bf k'}\langle n_{\bf k'}
    \rangle \omega_{\bf k'} ] \label{eq:magren} \\
    & = & \omega_{\bf k}^{0}[1 - \kappa {\mathcal E}_{tot}],
    \nonumber
\end{eqnarray}
where ${\mathcal E}_{tot}$ is the total energy of the excited SWs
\cite{ref:Mattis1988}.  Similar, more complicated expressions hold for
antiferromagnets and for different crystal structures.  In our
experiments, the $X$-$M$ line shape reflects this renormalization, as
$\alpha(\omega, \{ n_{\bf k} \})\simeq\rho_{e-m}(\omega, \{n_{\bf k}
\})$ now depends on the time dependent distribution of photo-excited
SWs, and the pump-probe experiment measures the time evolution of
$\sum_{\bf k}\frac{d\alpha}{dn_{\bf k}}n_{\bf k}(t)$.  In principle,
the exciton dispersion relation should also be renormalized for the
same reasons, but the exciton bandwidth is a factor of 2 narrower than
that of the SWs, and its dispersion is weakest near the zone boundary,
so this effect contributes little to the overall line shape change.
Numerical simulations confirm these arguments. Since $\alpha$ is
proportional to the integrated JDOS given in Eq.~(\ref{eq:jdos}), the
total derivative $\frac{d\alpha}{dn_{\bf k}}$ includes two distinct
contributions, one from the level shifts $\frac{d\omega_{\bf
    k}}{dn_{\bf k}}$ and the other from the change in the integration
volume associated with the level shifts.  Qualitatively, the level
shifts produce an overall redshift in the energy, while the change in
the integration volume reduces the spectral weight.

\begin{figure}[tbp]
     \includegraphics[width=3in]{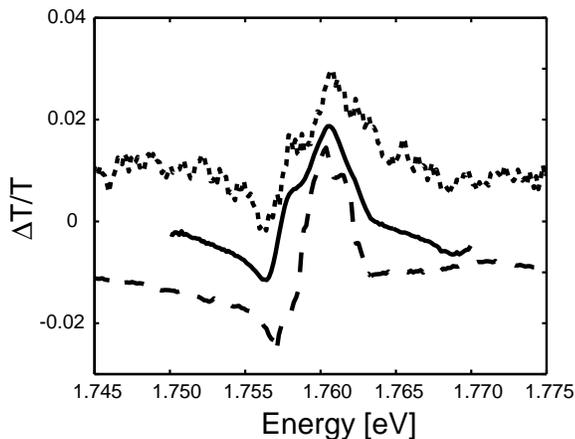}
\caption{Comparison of the saturated pump-probe line shape with
  simulated and actual changes due to a thermalized energy
  density. Solid line: pump probe line shape, $\times 3$; dotted line:
  thermal difference spectrum; dashed line: calculation using
  eqs. (\ref{eq:jdos}) and (\ref{eq:magren}).}
\label{fig:rencomp}
\end{figure}

It is interesting to compare the pump-probe signal at long times to
the change in the linear absorption induced by raising the
temperature. The SW energy absorbed from the laser is only $\sim2$\%
of the total absorbed energy, so from the total incident energy
density of 5.6 J/$\rm cm^3$ only $\sim100\ \rm{mJ/cm^3}$ is absorbed
by the SW system. From the known SW dispersion relation measured by
neutrons \cite{ref:Samuelson1969}, we may calculate the SW
contribution to the specific heat, $C_p\simeq C_v =
\frac{2\pi}{15}\frac{k_{B}^4}{D^3}T^3$, where $D \simeq 1.25 \times
10^{-28}\,{\rm J/cm}$. The energy density of one laser pulse thus
corresponds to a temperature change of $\sim30$ K in the magnetic
system. In Fig.~\ref{fig:rencomp}, we compare the saturated pump-probe
line shape to that expected from a 30 K temperature change,
based on the measured temperature dependence of the linear absorption.
For additional comparison, we also show the line shape calculated from
equations (\ref{eq:jdos}) and (\ref{eq:magren}), using a scale factor
$\kappa$ derived from the temperature dependent absorption spectra.
The pump-probe spectrum has been scaled up by a factor of three, which
may reasonably be attributed to the simplifications of
Eq.~(\ref{eq:magren}).

The agreement among these curves confirms our assignment of the
line shape to SWR, and shows that the energy density in the magnetic
system at long time delay is comparable to that absorbed directly by
the spin system from the pump beam.  The magnetic excitons absorb the
majority of the laser energy, serving as a reservoir.  The excess
energy is transferred rapidly to quenching sites, whereupon it is
dissipated over 100 ns to several microseconds via nonradiative
processes~\cite{ref:Henderson1989}.  In the steady state, a large
number of these defect states will be excited, together with the
steady-state phonon and SW distributions.  As a test for indirect
energy transfer to the magnetic system via defects and phonons, we
have performed two-color pump-probe experiments using a Coherent RegA
9000 regenerative amplifier system to create high intensity pump
pulses at 1.5 eV, well away from the $X$-$M$ absorption feature.  We
focused a portion of this beam onto a 3 mm sapphire crystal to
generate a white light continuum probe, and used an interference
filter to select a 15 meV spectral range spanning the $X$-$M$
absorption.  For absorbed pulse energy densities ranging from 1-100
times those used in the degenerate pump-probe experiments, we observed
no measureable change in the $X$-$M$ absorption, indicating that the
mechanisms for energy transfer from defect absorption into the
magnetic system occurs on time scales much longer than those of
interest here.  We conclude that the magnetic system behaves as a
quasi-closed system at least during the first nanosecond, and that the
dynamics which we observe is related to an intrinsic internal
thermalization of the optically induced, nonequilibrium SW population.

\begin{figure}[tbp]
     \includegraphics[width=3in]{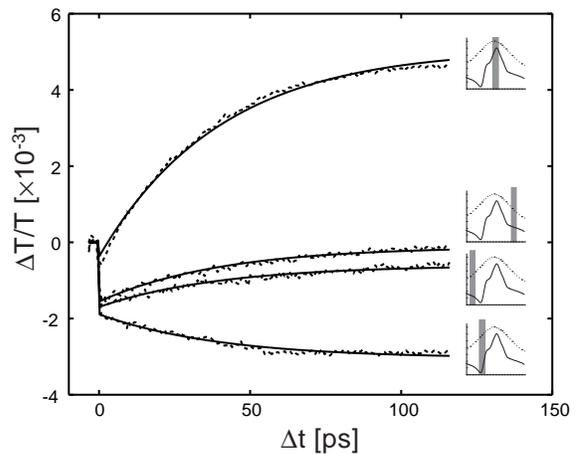}
\caption{Time-resolved pump-probe signal at different wavelengths over
  energy bandwidth $\hbar \Delta \omega = 20$ meV. Probe energy ranges
  are indicated in the insets accompanying each curve, and correspond
  to the following average energies, listed from top to bottom: 1.764
  eV, 1.771 eV, 1.754 eV, and 1.759 eV.}
\label{fig:pptrslv}
\end{figure}

We show the time evolution directly in Fig.~\ref{fig:pptrslv}, where
we plot the response at different wavelengths, keeping laser center
wavelength fixed.
As in Fig.~\ref{fig:specrslv}, we have subtracted a small background
component which is present at negative time delay, which is due to the
steady state heating of the sample. When the probe frequency is
outside the $X$-$M$-line $\Delta{\rm T/T}$ exhibits prompt PIA with a
decay time longer than 500 ps, not shown here. When the probe
frequency is inside the $X$-$M$-line $\Delta{\rm T/T}$ exhibits both
prompt PIA and picosecond dynamics.  The initial distribution of SWs
created by the laser are weighted heavily toward the zone boundary,
because of the factor of $4\pi k^{2}$ in the JDOS integral given in
Eq.~(\ref{eq:jdos}).  If all SWs contributed equally to the
renormalization, as suggested by Eq.~(\ref{eq:magren}), one would
expect the SWR and hence the $X$-$M$ line shape to undergo an abrupt
change at $\Delta t = 0$, and remain unchanged during the internal
thermalization of the SW system.  This clearly is not the case: the
initial population of $k \approx \pi/a$ SWs contributes little to the
$X$-$M$ renormalization line shape.

The exciton and magnon are initially created on neighboring sites and
should interact moderately with each other, but the difference in
their relative group velocities of $\sim$10 \AA/ps indicates that they
should be well separated after a picosecond or less, so we do not
believe that the observed change is associated with the decay of the
$X$-$M$ composite.  It is also possible that the process of optical
absorption in the presence of large $k$ excitations is not described
well by the particular approximation used here, but must include
additional many-particle interactions.  Such effects, however, would
need to suppress the contribution of SWR to the pump-probe line shape
by an order of magnitude.  The simplest explanation for our result is
that occupation at large $k$ produces weaker overall SWR than those at
the zone center.  Such strong $k$ dependence in SW interactions has
long been indicated theoretically, and the variation of the
interaction at short wavelengths may be so strong that the overall
interaction effects cancel~\cite{ref:Silberglitt1967}.

Regardless of the underlying reason for the difference between zone
center and zone boundary SW occupation, we can describe the dynamical
response phenomenologically by dividing the occupied SW states into
two different populations, those at the zone boundary ({\em b}) and
those at the zone center ({\em c}), with a boundary in reciprocal
space chosen to reproduce the experiments.  We assume that the decay
of the initial nonequilibrium energy density ${\mathcal E}_b =
\sum_{{\bf k} \in {\bf k}_{b}} n_{\bf k} \epsilon_{\bf k}$ into the
thermalized energy density ${\mathcal E}_c = \sum_{{\bf k} \in {\bf
    k}_{c}} n_{\bf k} \epsilon_{\bf k}$ is governed by a single
thermalization time, $\tau$, and the decay over long times is set by
an overall energy decay time, $\mathcal T$. Clearly, $\tau$ provides a
measure of the interactions coupling the zone boundary spin-waves to
those at lower energy, both directly and via phonons.  This process
may be described by the following phenomenological rate equations:
\begin{eqnarray}
\frac{d{\mathcal E}_b}{dt} & = & -{\mathcal E}_b/\tau - {\mathcal
  E}_b/{\mathcal T},\\
\frac{d{\mathcal E}_c}{dt} & = & {\mathcal E}_b/\tau - {\mathcal
  E}_c/{\mathcal T}. \nonumber
\end{eqnarray}

If we further assume that only the energy density due to zone center
SWs ${\mathcal E}_{c}$ is involved in the renormalization line shape,
we obtain the following equation for the time-dependent pump-probe
signal, valid for each energy within the $X$-$M$ absorption region:
\begin{equation}
    \Delta{\rm T/T} = a_{1}\Theta(t) + a_{2}[1 -
    \exp(-t/\tau)]\exp(-t/{\mathcal T}),
    \label{eq:trmodel}
\end{equation}
where $a_{1}$ and $a_{2}$ are prefactors which depend on the spectral
region of interest. The step function $\Theta(t)$ is required to
account for the PIA, which we have taken to be time independent,
consistent with experiments away from the $X$-$M$ peak. The weak
structure at early times, due to SWR, is also included in $a_{1}$,
though in principle this may be accounted for by an additional term.
We have divided the 10 nm spectral range given by our laser spectrum
into ten ranges of equal width, separated by 1 nm, and measured the
temporal pump-probe response in each range. We then found the best
global fits to the data of Eq.~(\ref{eq:trmodel}), in which $a_{1}$
and $a_{2}$ are allowed to vary with probe wavelength, and $\tau = 40
\pm 10 $ ps and ${\mathcal T} = 2 \pm 1$ ns are constrained to be the
same for all ten wavelengths.  The results for four of them are shown
by the theoretical curves in Fig.~\ref{fig:pptrslv}.  This simple
model captures very well the observed dynamics.

In summary, we have generated a macroscopic, nonequilibrium population
of SWs and observed its dynamics, using a pulsed laser spectroscopy.
At long times, the change in the absorption line shape is well
understood by assuming that the photogenerated SWs induce a
renormalization of the SW dispersion relation. At short times, our
results deviate sharply from the predictions of this model, indicating
that short wavelength SWs do not contribute to the renormalization
line shape.  The SWs form a quasi-closed system over 100 ps time
scales.  The evolution of this nonequilibrium population is consistent
with a simple model of SW thermalization, and the thermalization time
characterizes intrinsic spin-wave coupling. The effects reported here
provide us with information on elementary magnetic excitations that is
inaccessible through conventional techniques, which typically probe
only thermally occupied magnetic excitations.  Moreover, the technique
may be used quite generally in magnetic insulators, and may be applied
to a wide variety of optical magnetic excitations, including
two-magnon excitations and the recently assigned phonon-bimagnon
feature in the undoped cuprates~\cite{ref:Lorenzana1995}.  By using
multiple wavelength laser sources, this technique may be used in
conjunction with excitations across the Mott gap to probe directly the
interactions between charge and spin excitations in magnetic solids.

We would like to acknowledge L.~Sham for critical reading of the
manuscript. This work was supported by the Director, Office of Energy
Research, Office of Basic Energy Sciences, Division of Materials
Sciences of the U.~S.~Department of Energy under Contract No.
DE-AC03-76SF00098, and by the Stanford NSF-MRSEC Program. A.~B.~S.
acknowledges support by the German National Merit Foundation.

\end{document}